%% file: BELLE2-CONF-DRAFT-2022-017.tex
\documentclass[preprint,prd,tightenlines,superscriptaddress, nofootinbib]{revtex4-1}

\usepackage{graphicx}
\usepackage{dcolumn} 
\usepackage{colordvi}
\usepackage{color}
\usepackage{epstopdf}
\usepackage{amssymb}
\usepackage{amsmath}
\usepackage{url}
\graphicspath{{ps}}
\usepackage{hyperref}
\usepackage{tabularx}
\usepackage[columnwise,pagewise]{lineno}
\usepackage{siunitx}
\usepackage{pgf}
\usepackage{pgffor}
\usepackage{pgfplots}
\usepackage{tikz}
\usepackage{units}
\usepackage[super]{nth}
\usepackage{hepnames}
\usepackage{amssymb}
\usepackage{amsfonts}
\usepackage{upgreek}
\usepackage[tikz]{bclogo}
\usepackage{import}
\usepackage{multirow}
\usepackage{subcaption}
\input{definitions}
\usepackage{relsize}

\usetikzlibrary{pgfplots.colorbrewer,pgfplots.fillbetween}
\usetikzlibrary{shadows.blur}
\usetikzlibrary{shapes.symbols}
\usetikzlibrary{calc,intersections,matrix}
\usetikzlibrary{arrows,fadings,automata,snakes,shapes,shapes.misc,shapes.arrows,trees,positioning,calc,decorations.pathreplacing,arrows.meta}
\tikzfading[name=arrowfading, top color=transparent!0, bottom color=transparent!95]
\tikzset{arrowfill/.style={top color=blue!50, bottom color=blue,}}
\tikzset{>={Latex[width=1.5mm,length=1.5mm]}}
\tikzset{arrowstyle/.style={draw=black,arrowfill, single arrow,minimum height=#1, single arrow,
single arrow head extend=.4cm,}}

\tikzset{
    box/.style={
      rectangle,
	  color=#1,
      draw=black,
      fill=#1,
      thick,
      text=black,
      align=center,
      rounded corners=6pt,
      minimum height=1.5em
    }, 
    hbox/.style={
      rectangle,
      draw=black,
      fill=black,
      thick,
      text=white,
      align=center,
      rounded corners=6pt,
      minimum height=1.5em
    }, 
}
\colorlet{tree0}{blue}
\colorlet{tree100}{white!20!blue}
\colorlet{tree200}{white!40!blue}
\colorlet{tree300}{white!60!blue}
\colorlet{tree400}{white!80!blue}
\colorlet{tree500}{white}
\colorlet{tree600}{white!80!orange}
\colorlet{tree700}{white!60!orange}
\colorlet{tree800}{white!40!orange}
\colorlet{tree900}{white!20!orange}
\colorlet{tree1000}{orange}

\definecolor{Tblue}{HTML}{3465A4}
\definecolor{Tbluedark}{HTML}{204A87}
\definecolor{Tbluelight}{HTML}{729FCF}
\definecolor{Tbluelighter}{HTML}{8CC4FF}	
\definecolor{Tbrown}{HTML}{C17D11}	
\definecolor{Tbrowndark}{HTML}{8F5902}	
\definecolor{Tbrownlight}{HTML}{E9B96E}	
\definecolor{Tgray}{HTML}{888A85}
\definecolor{Tgraydark}{HTML}{555753}	
\definecolor{Tgraydarker}{HTML}{2E3436}	
\definecolor{Tgraylight}{HTML}{BABDB6}	
\definecolor{Tgraylight2}{HTML}{E4E6E2}	
\definecolor{Tgraylight3}{HTML}{F0F2EE}	
\definecolor{Tgreen}{HTML}{73D216}	
\definecolor{Tgreendark}{HTML}{4E9A06}
\definecolor{Tgreenlight}{HTML}{8AE234}	
\definecolor{Tred}{HTML}{CC0000}	
\definecolor{Treddark}{HTML}{A40000}
\definecolor{Tredlight}{HTML}{EF2929}
\definecolor{Tlilac}{HTML}{75507B}
\definecolor{Tlilacdark}{HTML}{5C3566}
\definecolor{Tlilaclight}{HTML}{AD7FA8}
\definecolor{Tyellow}{HTML}{EDD400}	
\definecolor{Tyellowdark}{HTML}{C4A000}	
\definecolor{Tyellowlight}{HTML}{FCE94F}
\definecolor{Torange}{HTML}{F57900}	
\definecolor{Torangedark}{HTML}{CE5C00}
\definecolor{Torangelight}{HTML}{FCAF3E}

\makeatletter
\let\LN@align\align
\let\LN@endalign\endalign
\renewcommand{\align}{\linenomath\LN@align}
\renewcommand{\endalign}{\LN@endalign\endlinenomath}
\let\LN@gather\gather
\let\LN@endgather\endgather
\renewcommand{\gather}{\linenomath\LN@gather}
\renewcommand{\endgather}{\LN@endgather\endlinenomath}
\makeatother

{}
\begin{document}

\vspace*{-3\baselineskip}
\resizebox{!}{3cm}{\includegraphics{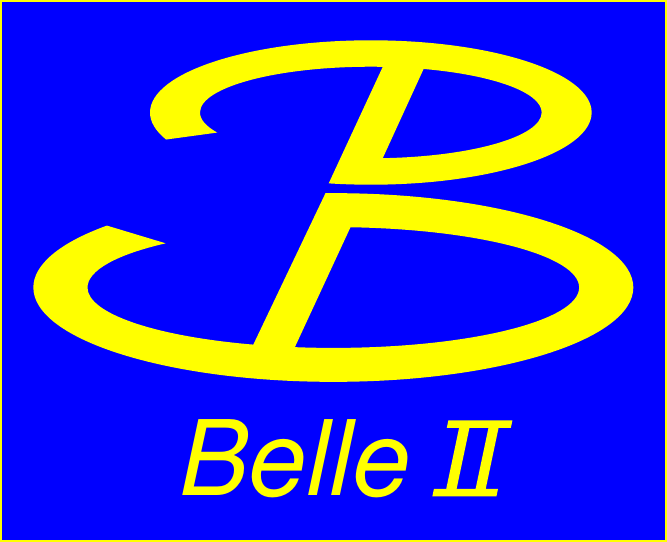}}

\vspace*{-5\baselineskip}
\begin{flushright}
BELLE2-CONF-PH-2022-017\\
November 7, 2022
\end{flushright}

\title{ \quad\\[0.5cm] Determination of $|V_{ub}|$ from untagged $B^0\to\pi^- \ell^+ \nu_{\ell}$ decays using 2019-2021 Belle~II data}
\input{authors-conf2022-plain}

\collaboration{The Belle~II Collaboration}
\noaffiliation

\begin{abstract}
\input{abstract}
\end{abstract}

\pacs{\input{pacs}}
\maketitle

\input{main}

\newpage
\section{ACKNOWLEDGEMENTS}
\input{acknowledgements}

\bibliography{references}
\bibliographystyle{unsrt}

\end{document}

%% file: definitions.tex
\newcommand*{\Btopilnu}{\ensuremath{B^0\to\pi^- \ell^+ \nu_{\ell}}}
\newcommand*{\Btopienu}{\ensuremath{B^0\to\pi^- e^+ \nu_{e}}}
\newcommand*{\Btopimunu}{\ensuremath{B^0\to\pi^- \mu^+ \nu_{\mu}}}
\newcommand*{\Bztopiellnu}{\ensuremath{B^0\to\pi^- \ell^+ \nu_{\ell}}}

\newcommand*{\BtoXulnu}{\ensuremath{B\to X_{u} \ell\nu}}
\newcommand*{\BtoXclnu}{\ensuremath{B\to X_{c} \ell\nu}}
\newcommand*{\BtoXlnu}{\ensuremath{B\to X \ell\nu}}

\newcommand*{\Btopiellnu}{\ensuremath{B\to\pi\ell \nu}}
\newcommand*{\Btorholnu}{\ensuremath{B\to \rho \ell\nu}}
\newcommand*{\BtoDlnu}{\ensuremath{B\to D \ell\nu}}
\newcommand*{\BtoDstlnu}{\ensuremath{B\to D^{*} \ell\nu}}

\newcommand*{\thetamiss}{\ensuremath{\theta_{\mathrm{miss}}}}
\newcommand*{\cosThetaBY}{\ensuremath{\cos \theta_{BY}}}

\newcommand*{\deltae}{\ensuremath{\Delta E}}
\newcommand*{\mbc}{\ensuremath{M_{bc}}}

%% file: authors-conf2022-plain.tex
\collaboration{The Belle II Collaboration}
\author{F. Abudin{\'e}n}
\author{I. Adachi}
  \author{K. Adamczyk}
  \author{L. Aggarwal}
  \author{P. Ahlburg}
  \author{H. Ahmed}
  \author{J. K. Ahn}
  \author{H. Aihara}
  \author{N. Akopov}
  \author{A. Aloisio}
% ...

%% file: abstract.tex
We present an analysis of the charmless semileptonic decay $B^0\to\pi^- \ell^+ \nu_{\ell}$, where $\ell = e, \mu$, from 198.0 million pairs of $B\bar{B}$ mesons recorded by the Belle~II detector at the SuperKEKB electron-positron collider.
The decay is reconstructed without identifying the partner $B$ meson.
The partial branching fractions are measured independently for $B^0\to\pi^- e^+ \nu_{e}$ and $B^0\to\pi^- \mu^+ \nu_{\mu}$ as functions of $q^{2}$ (momentum transfer squared), using 3896 $B^0\to\pi^- e^+ \nu_{e}$ and 5466 $B^0\to\pi^- \mu^+ \nu_{\mu}$ decays.
The total branching fraction is found to be $(1.426 \pm 0.056 \pm 0.125) \times 10^{-4}$ for $B^0\to\pi^- \ell^+ \nu_{\ell}$ decays, where the uncertainties are statistical and systematic, respectively.
By fitting the measured partial branching fractions as functions of $q^{2}$, together with constraints on the nonperturbative hadronic contribution from lattice QCD calculations, the magnitude of the Cabibbo-Kobayashi-Maskawa matrix element $V_{ub}$, $(3.55 \pm 0.12 \pm 0.13 \pm 0.17) \times 10^{-3}$, is extracted.
Here, the first uncertainty is statistical, the second is systematic and the third is theoretical.

%% file: main.tex
\section{Introduction}

The Cabibbo-Kobayashi-Maskawa (CKM) matrix elements are fundamental parameters of the Standard Model (SM) of particle physics ~\cite{Kobayashi:1973fv}.
The magnitude of the matrix element $V_{ub}$ can be determined by measuring the differential decay rate of $B\rightarrow\ X_{u}\ell\nu$ events, which is proportional to $|V_{ub}|^{2}$.
Here, $X_{u}$ is a charmless hadronic final state and $\ell$ is a light charged lepton.
One method to measure $|V_{ub}|$ is inclusive, where no specific $X_{u}$ final state is reconstructed, but rather the sum of all possible final states is analysed.
This is in contrast to the exclusive method in which a specific final state is studied. 
The two methods have complementary uncertainties introduced by theoretical QCD descriptions.
In inclusive decays, these descriptions involve the calculation of the total semileptonic rate, while in the exclusive case they take the form of parameterizations of the low-energy strong interactions (form factors).
However, the results for $|V_{ub}|$ obtained by the two methods differ significantly~\cite{Amhis:2019ckw}. 
In order to resolve this disagreement, further measurements using novel approaches are beneficial.

The decay \Btopilnu{} (with charge conjugation implied throughout) is experimentally and theoretically the most reliable mode to measure $|V_{ub}|$ through an exclusive channel at the $B$ Factories.
Here, we present a study of this decay using data from the Belle~II detector located at the SuperKEKB electron-positron collider at KEK, in Japan.
The signal decay is reconstructed from decays of $B^{0}$ mesons in $e^{+}e^{-}\rightarrow \Upsilon(4S)\rightarrow B^{0}\overline{B^{0}}$ events.
The reconstruction method employed here is called untagged since the signal lepton and pion candidates are selected without prior reconstruction (tagging) of the partner $B$ meson.
This leads to a high signal efficiency, but also a low purity due to higher combinatorial background from the partner $B$ meson and increased backgrounds from $e^{+}e^{-}$ collisions that produce light quark pair (continuum) events.

The events are separated into six disjoint intervals (bins) of squared momentum transfer from the $B$ meson to the pion, $q^{2}$.
After selecting signal events and suppressing backgrounds, the signal yields are extracted from an extended likelihood fit to the binned two-dimensional distribution of the energy difference \deltae{}, and the beam-constrained mass \mbc{}, both defined below, in bins of $q^{2}$. 
Theoretical form-factor predictions from lattice QCD are combined with the measured differential branching fractions to determine $|V_{ub}|$~\cite{FermilabMILC}.

\section{Detector, data set and simulation}
\subsection{Belle~II detector}

The Belle~II detector~\cite{Belle-II:2010dht} is located at the SuperKEKB~\cite{Akai:2018mbz} asymmetric-energy electron-positron collider, running at or slightly below the $\Upsilon(4S)$ resonance energy.
The detector is composed of subdetectors arranged around the collision point in a cylindrical geometry. 
The longitudinal direction is defined by the $z$-axis, which points approximately along the electron-beam direction, while the transverse plane is defined by the $x$- and $y$-axes, which point radially out of the detector ring.
A superconducting magnet producing a uniform 1.5~T magnetic field oriented along the detector axis is installed outside all but the last subdetector.
Particles produced at the interaction point first traverse the vertex detector composed of a two-layer pixel detector and a four-layer silicon strip detector that measure positions of decays into charged particles.
Currently, however, the second layer of the pixel detector covers only one sixth of the azimuthal angle.

Trajectories of charged particles (tracks) are further determined from a central drift chamber (CDC) that has a polar angle acceptance of 17--150$^{\circ}$. 
We use two Cherenkov particle-identification subdetectors: the time-of-propagation counter is located around the barrel region of the detector and the aerogel ring-imaging Cherenkov detector is located in the forward endcap. 
Photon- and electron-energy measurements in the barrel region and the endcaps are performed using the electromagnetic calorimeter, which is composed of CsI(Tl) crystals. 
Surrounding the magnet is the resistive-plate-based $K^{0}_{L}$ and muon detector, which also doubles as a flux return for the magnet. 

Using specific-ionization energy-loss information provided by the CDC and information from the two particle-identification subdetectors and the $K^{0}_{L}$ and muon detector, charged particles of different masses are distinguished and particle-identification (ID) variables are constructed. 
These are normalized ratios of likelihoods for one charged-particle hypothesis to the sum of all possible charged-particle likelihoods. 

The detector performance is monitored using control samples; we correct for time-dependent changes such as shifts in the magnetic field map or the photon energy scale.
In addition, the efficiencies and misidentification probabilities of pions, kaons, electrons and muons are monitored separately by charge in bins of particle momentum and polar angle using well-known physics control modes, such as $J/\psi \to \ell^{+}\ell^{-}$ and $D^{*}\rightarrow D^{0}[K\pi]\pi$.

\subsection{Data set}

The primary data set used in this analysis is collected at a center-of-mass (CM) energy of $\sqrt{s} = $ 10.58~GeV, corresponding to the mass of the $\Upsilon$(4S) resonance. 
A data set corresponding to an integrated luminosity of $189$~fb$^{-1}$ is collected at this CM energy, equivalent to a sample of ($198.0\pm 3.0$) million $\Upsilon(4S)\rightarrow B\overline{B}$ events.
In addition, we use a sample corresponding to $18$~fb$^{-1}$ of off-resonance collision data, collected at a CM energy 60~MeV below the $\Upsilon$(4S) resonance, to describe background from continuum processes.
These include $q\overline{q}$ continuum background, i.e., $e^+ e^- \to u\bar u,\, d\bar d,\, s\bar s$ and $c\bar c$.
The off-resonance data also contain other continuum backgrounds, such as $e^+ e^- \to\tau^+\tau^-$, and two-photon processes, where $e^+ e^- \to e^+e^-\ell^+\ell^-$.

\subsection{Monte Carlo simulation}

Simulated Monte Carlo (MC) samples of continuum and $\Upsilon(4S)\rightarrow B\overline{B}$ events, corresponding to an integrated luminosity of 1~ab$^{-1}$ are used.
We use these to identify efficient background-discriminating variables and to form fit templates for signal extraction. 
In addition, MC samples of signal \Btopilnu{} decays are used to obtain the reconstruction efficiencies and study the key kinematic distributions.
In order to describe the composition of the $B\to X_u \ell \nu$ decays, 50 million events each of resonant and nonresonant $B^0\to X_u \ell \nu$, and $B^{\pm}\to X_u \ell \nu$ are simulated.

The $B\bar{B}$ events are generated using the event generator \texttt{EvtGen}~\cite{Lange:2001uf}, while the $q\overline{q}$ continuum events are generated using \texttt{PYTHIA}~\cite{Sjostrand:2014zea}.
Tau-pair events are generated using \texttt{KKMC}~\cite{Jadach:1999vf}, and their decays are handled using \texttt{TAUOLA}~\cite{Jadach:1990mz}.
The simulation of two-photon processes is performed using \texttt{AAFH}~\cite{BERENDS200282}.
Backgrounds induced by the beams are mixed into the MC samples using simulations of beam-induced backgrounds~\cite{BeamBKG}.
The propagation of the particles through the detector and the resulting interactions are simulated using \texttt{Geant4}~\cite{Agostinelli:2002hh} and final-state radiation is modeled using \texttt{PHOTOS}~\cite{Barberio:1990ms}. 
All recorded and simulated events are handled with the analysis software framework \textit{basf2}~\cite{Kuhr:2018lps}.

The branching-fraction values of the \BtoXlnu{} decays assumed in the simulation are obtained from current world averages~\cite{Zyla:2020zbs}, in combination with assumed isospin symmetry, following the procedure in Ref.~\cite{Bernlochner:2016bci} for \BtoXclnu{} decays. 
The remaining difference between the sum of the exclusive \BtoXclnu{} decay branching fractions and the measured total branching fraction accounts for approximately 4\% of the \BtoXclnu{} decays and is assumed to be saturated by $B\rightarrow D^{(*)}\eta \ell\nu$ decays.
For \BtoXulnu{} decays, the difference between the sum of the exclusive branching fractions and the measured total branching fraction is assumed to be saturated by nonresonant \BtoXulnu{} decays with multiple pions in the final state following Ref.~\cite{Lange:2005yw}.
The nonresonant \BtoXulnu{} decays account for approximately 80\% of the \BtoXulnu{} decays.

The nonresonant \BtoXulnu{} composition is described by implementing a hybrid model ~\cite{PhysRevD.41.1496}, following closely the method in Ref.~\cite{Belle:2019iji}.
This approach combines the exclusive and nonresonant decay rates in bins of $X_{u}$ particle mass $m_{X}$, the rest-frame lepton energy $E_{\ell}$, and $q^{2}$, in order to recover the inclusive rates. 

For the form factors of \BtoDlnu{} and \BtoDstlnu{} decays, we use the Boyd-Grinstein-Lebed parameterization~\cite{PhysRevLett.74.4603} with central values from Ref.~\cite{PhysRevD.93.032006} and \cite{Ferlewicz:2020lxm}, respectively.
For the $B \rightarrow \rho\ell\nu$ and $B \rightarrow \omega\ell\nu$ form factors we choose the Bourrely-Caprini-Lellouch (BCL) parameterization~\cite{PhysRevD.79.013008} with central values from Ref.~\cite{Bernlochner:2021rel}.
Finally, for the form-factor description of the $B\rightarrow \eta^{(')} \ell\nu$ decays a light-cone sum rule calculation is chosen~\cite{Duplancic:2015zna}.

\section{Event reconstruction and selection}

\subsection{Event reconstruction}
We begin signal reconstruction by identifying track candidates that pass certain quality criteria. 
The extrapolated trajectories must pass within 1~cm (3~cm) of the interaction point transverse (parallel) to the beam.
Furthermore, charged particles must have a transverse momentum greater than 0.05~GeV (we use natural units with $c=1$ throughout) and polar angles within the CDC acceptance, in order to suppress misreconstructed track candidates.
We discard events with fewer than five tracks satisfying the above criteria.

In the remaining events, we select lepton candidates from among the selected tracks by requiring their CM momentum $p^{*}_{\ell}$ to satisfy 1.0~$<p^{*}_{\ell} <3.2$~GeV.
Electron and muon candidates are required to have an electron or muon likelihood greater than 0.9, respectively.
The average electron (muon) efficiency is 91 (93)\%.
The hadron misidentification rates are 0.2\% for the electron and 3.3\% for the muon selection, respectively.
Electron-candidate four-momenta are corrected for bremsstrahlung by adding to them the four-momenta of photons identified within a cone around the electron direction.

Signal-pion candidates are selected from all remaining tracks that pass within 2~cm (4~cm) of the interaction point transverse (parallel) to the beam.
They are required to have a charge opposite to the lepton candidate and to have a pion likelihood greater than 0.1 in order to suppress misidentification of kaons as pions.
To improve the particle-identification performance, we require the pion candidates to leave at least 20 random measurement points (hits) in the CDC.
The resulting average pion efficiency is 84\% with a kaon misidentification rate of 7\%.

We reduce backgrounds by removing candidates with kinematic properties inconsistent with the signal $B$ decay. 
Under the assumption that only a single massless particle is not included in the event reconstruction, the angle between the $B$ meson and the combination of the lepton and pion candidates, denoted $Y$, is fully determined,
\begin{equation}	
\cos \theta_{BY} = \frac{2E_{B}E_{Y}-m_{B}^{2}-m_{Y}^{2}}{2p_{B}p_{Y}},
\label{eq:thetaBY}
\end{equation}
where $E_{Y}$, $p_{Y}$, and $m_{Y}$ are the energy, magnitude of the three-momentum, and invariant mass of the $Y$, respectively. The energy $E_{B}$ and the magnitude of the three-momentum of the $B$ meson $p_{B}$ are calculated from the beam properties, and $m_{B}$ is the mass of the $B$ meson~\cite{Zyla:2020zbs}. 
For correctly reconstructed signal decays, \cosThetaBY{} must lie between $-1$ and 1. 
However, to allow for a background-dominated region, we set a looser requirement of $|\cosThetaBY| < 2.2$.
The event shape discriminates between $B\overline{B}$ and continuum events.
Thus we require the second normalized Fox-Wolfram moment~\cite{FWM} to be less than 0.4.

We obtain the kinematic properties of the neutrino by assuming that the sum of the remaining tracks and electromagnetic energy-depositions (clusters) in the event, called the rest of event (ROE), represents the partner $B$ meson. 
From energy and momentum conservation, we construct a missing four-momentum in the CM frame, 
\begin{equation}	
(E^{*}_{\mathrm{miss}},\vec{p}^{*}_{\mathrm{miss}}) =  (E_{\Upsilon(4S)},\vec{p}_{\Upsilon(4S)})  - \left( \sum_{i} E^{*}_{i} , \sum_{i} \vec{p}^{*}_{i}\right),
\label{eq:p_miss}
\end{equation}
where $E^{*}_{i}$ and $\vec{p}^{*}_{i}$ correspond to the CM energy and momentum of the $i$th track or cluster in the event, respectively.
This yields the neutrino momentum, $\vec{p}^{*}_{\nu} = \vec{p}^{*}_{\mathrm{miss}}$, and energy,  $E^{*}_{\nu} = |\vec{p}^{*}_{\nu}| = |\vec{p}^{*}_{\mathrm{miss}}|$.

Since all reconstructed tracks and clusters contribute to the resolution of the neutrino momentum, obtaining a pure ROE is critical.
To reduce the impact of clusters from beam-induced backgrounds, acceptance losses, or other effects, we impose further criteria. 
We only consider clusters that are within the CDC acceptance and have transverse momenta in forward, barrel, and backward directions greater than 0.03, 0.04, and 0.06~GeV, respectively.
The transverse momentum of a cluster is defined using the energy and the location with respect to the interaction point of the energy deposition.
The clusters are required to be detected within 200~ns of the collision time, which is approximately five times the mean timing resolution of the calorimeter.
The clusters also have to consist of more than a single calorimeter crystal. 
In addition to removing background particles from the ROE, we must account for particles that may escape undetected.
To reduce the impact of these events, we require that the polar angle of the missing momentum in the laboratory frame $\theta_{\mathrm{miss}}$ is within the CDC acceptance.

\subsection{Signal extraction variables}

We calculate $q^2$ from $q^2=(p_B-p_{\pi})^2$, and thus need a way to estimate the $B$ momentum vector.
One existing method, called the \textit{Diamond Frame}~\cite{DiamondFrame}, takes the weighted average of four possible $\vec{p}_B$ vectors uniformly distributed in azimuthal angle on the cone defined by \cosThetaBY{} using weights of $\sin^2\theta_{B}$, which expresses the prior probability of the $B$ flight direction in $\Upsilon(4S)$ decays with respect to the beam axis.
A second method, called the \textit{ROE method}~\cite{Waheed:2019lxm}, assumes the signal $B$ vector to be the vector on the \cosThetaBY{} cone that is closest to antiparallel to the ROE momentum vector. 
We introduce a new method that combines these two by multiplying the Diamond Frame weights by $\frac{1}{2}(1-\vec{p}_B\cdot \vec{p}_{\mathrm{ROE}}/(|\vec{p}_B||\vec{p}_{\mathrm{ROE}}|))$ and averaging over ten vectors uniformly distributed on the cone.
We adopt this combined method because, in simulation, it assigns reconstructed signal candidates to the correct $q^2$ bin more often than other methods do, leading to a reduction in the bin migrations of up to 2\% and resolutions in $q^2$ ranging from 0.14--0.45~GeV$^{2}$.
We divide $B$ candidates into six $q^2$ bins with the following labels: $q1: q^2\in[0, 4]$, $q2: [4, 8]$, $q3: [8, 12]$, $q4: [12, 16]$, $q5: [16, 20]$, $q6: [20,\infty]$~GeV$^{2}$.

Using ROE information, two further variables that test the consistency of a candidate with a signal $B$ decay are the beam-constrained mass, defined as
\begin{equation}	
M_{bc} = \sqrt{E^{*2}_{\mathrm{beam}} -|\vec{p}^{\,*}_{B}|^{2}} =\sqrt{\left(\frac{\sqrt{s}}{2 }\right)^{2} -|\vec{p}^{\,*}_{B}|^{2}}
\label{eq:mbc}
\end{equation}
and the energy difference, defined as
\begin{equation}
\Delta E= E^{*}_{B} - E^{*}_{\mathrm{beam}} = E^{*}_{B} - \frac{\sqrt{s}}{2 },
\label{eq:deltae}
\end{equation} 
where $E^{*}_{\mathrm{beam}}$, $E^{*}_{B}$ and $\vec{p}^{\,*}_{B}$ are the single-beam energy, reconstructed $B$ energy, and reconstructed $B$ momentum all determined in the $\Upsilon(4S)$ rest frame, respectively. 
The reconstructed $B$ energy (momentum) is given by the sum of the reconstructed energies (momenta) of the signal lepton and pion and the inferred neutrino energy (momentum) described above.
We define a fit region in $\Delta E$ and $M_{bc}$, corresponding to $|\Delta E |<0.95$~GeV and $5.095 < M_{bc} < 5.295$~GeV.

\section{Background suppression}
\subsection{Background categories}
\label{subsec:categ}

The sample can be separated into two main categories: $B\overline{B}$ events and non-$B\overline{B}$ events. 
For the $B\overline{B}$ events we define a subcategory that combines signal and combinatorial signal events. 
In combinatorial signal either the pion or lepton candidate is incorrectly identified.
We further split the $B\overline{B}$ background into the two largest semileptonic backgrounds, $B\rightarrow X_{c} \ell\nu$ and $B\rightarrow X_{u} \ell\nu$. 
The remaining $B\overline{B}$ events are combined into an \emph{other} $B\overline{B}$ category, mainly composed of candidates with misidentified leptons or with leptons from secondary decays.
The non-$B\overline{B}$ events are combined into a continuum background category, which contains $q\overline{q}$ and other continuum events.

\subsection{Boosted decision trees}

In order to further reduce the $B\overline{B}$ and $q\overline{q}$ backgrounds we train boosted decision trees (BDTs) using the FastBDT methodology~\cite{FastBDT}, which differentiate between signal and each background category separately.
Since the background composition is different in each $q^{2}$ bin, we train the classifier and optimize the selection separately for each $q^{2}$ bin.
This results in a total of 12 BDTs. 
In each training we combine the electron and muon modes, since they have similar background compositions.
We use equal amounts of signal and background simulation events, corresponding to 200~fb$^{-1}$ of $q\overline{q}$ simulated data, and 600~fb$^{-1}$ of $B\overline{B}$ simulated data, due to the low $B\overline{B}$ background retention.

Two input variables, \cosThetaBY{} and \thetamiss{}, are common to the $q\overline{q}$ and $B\overline{B}$ suppression.
Further input variables in the $q\overline{q}$ suppression BDT are based on the thrust axis, which is the axis that maximizes the sum of the projected momenta of all charged particles in the event.
Distinct thrust axes can be defined for the signal $B$ and the ROE.
Their magnitudes, and the angle between the two axes $\cos\theta_{T}$ serve as input variables. 
In addition, three cones with opening angles of 10, 20, and 30$^{\circ}$ centered around the thrust axis are defined, and the momentum flow into each of the three cones is added as an input variable~\cite{CLEOCones}.
The input variables of the $B\overline{B}$ suppression BDTs are the number of tracks, the angle between the lepton and pion momenta, and the momentum of the ROE.
In addition, a vertex fit to the pion and lepton candidates is performed using TreeFit~\cite{TreeFit} and the $\chi^{2}$ probability is used as a discriminating feature.
The remaining two variables are the cosines of the angles between the signal $B$ momentum vector and the vector connecting its fitted vertex to the interaction point in the plane parallel and perpendicular to the beam axis, respectively.
The shared input variable \cosThetaBY{} is one of the most discriminating variables in both the suppression of continuum and $B\overline{B}$ events.
The $\cos\theta_{T}$ and the $\chi^{2}$ vertex fit probability also provide high discriminating power in the suppression of continuum and $B\overline{B}$ backgrounds, respectively.
We identify the optimal selection criterion on the output classifier within each $q^{2}$ bin by maximizing the ratio between the number of signal events and the square root of the sum of the number of signal and background events, as predicted by simulation.

After selecting on the output classifiers, the continuum background contains a significant amount of residual two-photon background, especially $ee\tau\tau$ events, in the electron mode.
Therefore, for the electron mode only, we train another BDT common to all $q^{2}$ bins, to separate signal from $ee\tau\tau$ events.
We use a sample corresponding to 2~ab$^{-1}$ of $ee\tau\tau$ simulated data and an equivalent amount of signal events.
The input variables include the total charge in the event, \thetamiss{}, the polar angle of the lepton, the number of clusters, the mass of the ROE, and $\xi_{z}$. 
Here, the variable $\xi_{z}$ is calculated as the sum of the momenta parallel to the beam axis divided by the sum of the energies of all charged particles in the laboratory frame.
None of the selected input variables are strongly correlated with \deltae{} or \mbc{}.

\subsection{Best candidate selection and efficiencies}
At this stage, an average of $1.02$ candidates remain per selected event.
In events with multiple candidates, we randomly select one and discard the rest
The overall signal efficiencies range from 8\% to 17\% in the electron mode and from 13\% to 23\% in the muon mode, depending on the $q^{2}$ bin.

\section{Signal extraction}
\subsection{Continuum treatment}

Because of the small off-resonance sample size, using the data directly as a fit template introduces large systematic uncertainties.
Instead, we weight the simulated continuum candidates in order to use the resulting template during signal extraction.
We therefore compare the $q^{2}$ shapes of simulated continuum data and off-resonance data. 
We observe a normalization difference in the electron mode, and to a lesser extent in the muon mode. 
After correcting the simulated continuum data for the total normalization, any residual difference in the $q^{2}$ spectrum is resolved by correcting the $q^{2}$ spectrum bin-by-bin.
This approach relies on the assumption that the difference between off-resonance data and the simulated continuum sample is independent of \deltae{} and \mbc{}.
Any observed deviation from this assumption is treated as a systematic uncertainty and explained in Section~\ref{sec:sys}.

\subsection{Fit setup}

\begin{figure*}[ht!]
	\begin{subfigure}[c]{0.495\textwidth}
		\centering
		\includegraphics[width=0.95\linewidth]{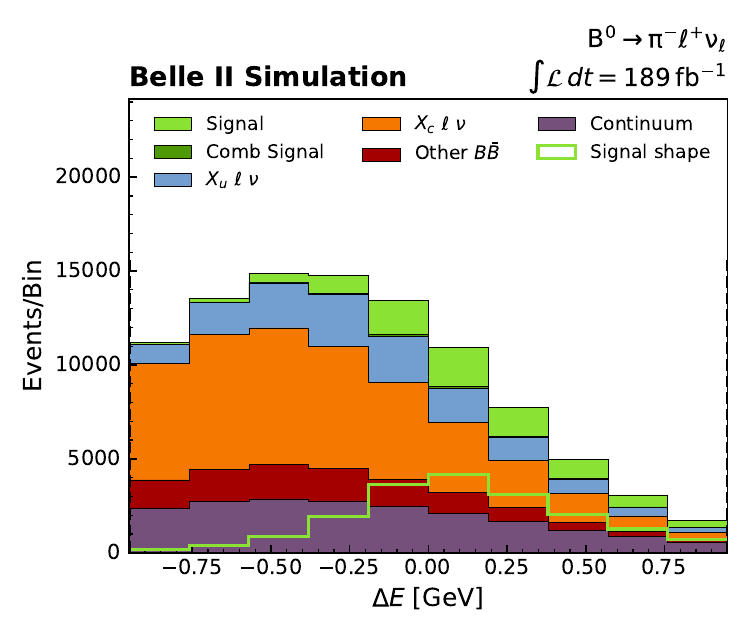}
	\end{subfigure}
	\begin{subfigure}[c]{0.495\textwidth}
		\centering
		\includegraphics[width=0.95\linewidth]{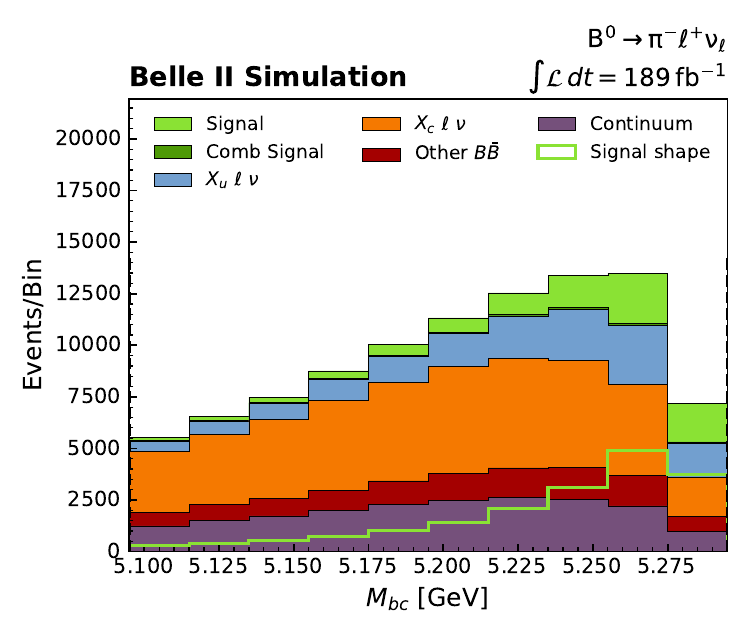}
	\end{subfigure}
	\caption{Simulated distributions of (left) \deltae{} and (right) \mbc{} integrated over the six $q^{2}$ bins for \Bztopiellnu{} decays.}
	\label{fig:deMbc_shapes}
\end{figure*}

We extract the signal by performing an extended likelihood fit to the binned two-dimensional distribution of \deltae{} and \mbc{} in bins of $q^{2}$.
In total we have 5 (\deltae{}) $\times$ 4 (\mbc{}) $\times$ 6 ($q^{2}$) $=$ 120 bins.
The distributions of \deltae{} and \mbc{} integrated over the six $q^{2}$ bins for \Bztopiellnu{} decays are shown in Figure~\ref{fig:deMbc_shapes}. 
The likelihood to be maximized is
\begin{equation}	
\mathcal{L}(s_{j},b_{k}) = \prod_{i} \mathrm{Poisson}(N_{i}|\sum_{j}s_{ij}+\sum_{k}b_{ik}),
\label{eq:likelihood}
\end{equation}
where $N_{i}$ is the observed number of events in bin $i$, $s_{ij}$ is the number of events in bin $i$ of signal fit template $j$, and $b_{ik}$ is the number of events in bin $i$ of background fit template $k$. 
The combinatorial signal is included in the signal yield. 
The backgrounds are split into four templates according to the description in Section~\ref{subsec:categ}.
In addition, there is one independent signal template for each of the six $q^{2}$ bins, leading to a total of ten templates.
The templates are constructed from 1~ab$^{-1}$ of simulated data, where we use the weighted simulated continuum events to extract the continuum template. 
We use a Gaussian penalty factor to constrain the continuum yield to the scaled off-resonance yield.

\subsection{Fit results}

\begin{figure*}[ht!]
	\begin{subfigure}[c]{0.495\textwidth}
		\centering
		\includegraphics[width=0.95\linewidth]{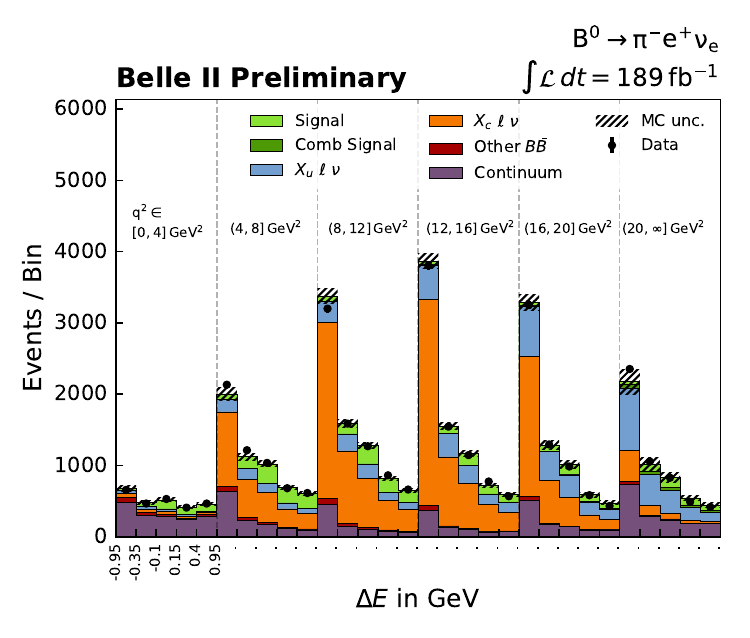}
	\end{subfigure}
	\begin{subfigure}[c]{0.495\textwidth}
		\centering
		\includegraphics[width=0.95\linewidth]{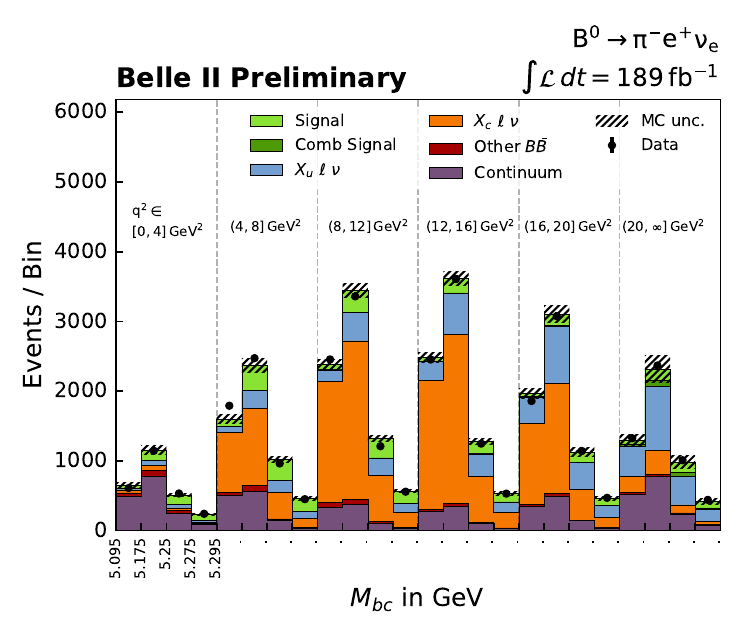}
	\end{subfigure}
	\begin{subfigure}[c]{0.495\textwidth}
		\centering
		\includegraphics[width=0.95\linewidth]{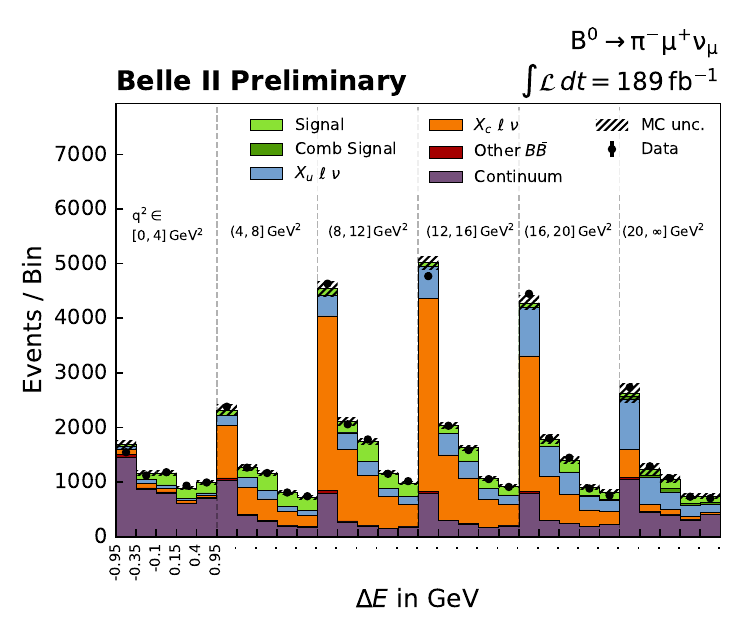}
	\end{subfigure}
	\begin{subfigure}[c]{0.495\textwidth}
		\centering
		\includegraphics[width=0.95\linewidth]{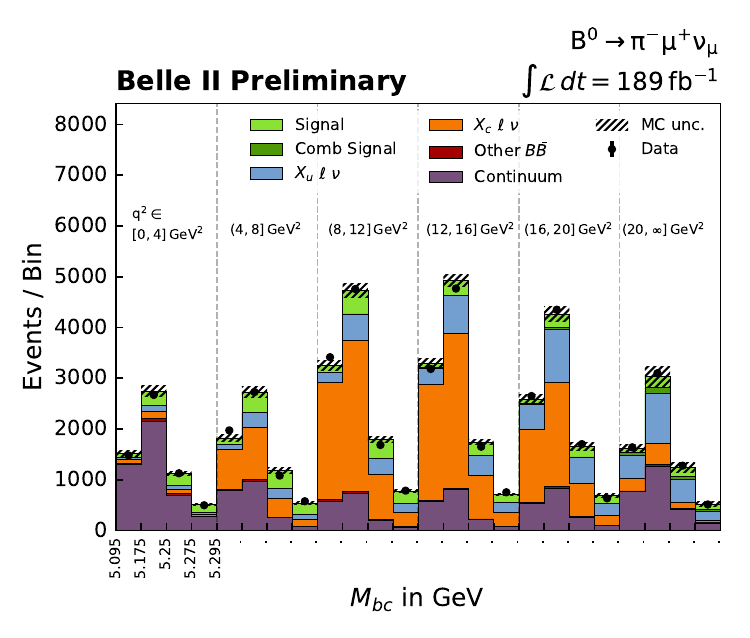}
	\end{subfigure}
	\caption{Distributions of (left) \deltae{} and (right) \mbc{} in the six $q^{2}$ bins for (top) \Btopienu{} and (bottom) \Btopimunu{} candidates reconstructed in Belle~II data with fit projections overlaid.
    }
	\label{fig:fitprojections}
\end{figure*}

The fit projections of \deltae{} and \mbc{} in each $q^{2}$ bin are shown in Figure~\ref{fig:fitprojections}.
The correlations between the component yields are all lower than 0.7.
The highest observed correlations occur between the \BtoXclnu{} and continuum yields.
In the higher $q^{2}$ bins, the signal becomes increasingly correlated to the \BtoXulnu{} yield.

The $q^{2}$ bin migrations in signal range from 3.7--9.8\%.
For correctly reconstructed signal they are largest in the lowest $q^{2}$ bin and decrease at higher $q^{2}$.
The inclusion of combinatorial signal, however, also results in larger bin migrations in the highest $q^{2}$ bin.
We correct the signal yields obtained from the fit for the migrations by applying the inverse detector response matrix~\cite{cowan1998statistical}. 
The corrected yields are given in Table~\ref{tab:sig_yields}. 
The \Btopienu{} signal yields are lower than the \Btopimunu{} signal yields.
This can partly be attributed to lower signal efficiencies due to the additional selection on the two-photon background suppression BDT.

\begin{table}[ht!]
	\begin{center}
	    \caption{Signal yields corrected for bin migrations in each $q^{2}$ bin with statistical and systematic uncertainties. The boundaries of the $q^{2}$ bins are given in the text above.}
		\begin{tabular}{ccc}
		    \hline
		    \hline
			$q^{2}$ bin & \multicolumn{1}{c}{\Btopienu} & \multicolumn{1}{c}{\Btopimunu} \\
			\hline
			$q1$ & 426 $\pm$ 49 $\pm$ 73 & 749 $\pm$ 74 $\pm$ 400\\
			$q2$ & 927 $\pm$ 56 $\pm$ 133 & 1076 $\pm$ 61 $\pm$ 164\\
			$q3$ & 856 $\pm$ 64 $\pm$ 96 & 1238 $\pm$ 77 $\pm$ 128\\
			$q4$ & 577 $\pm$ 64 $\pm$ 64 & 819 $\pm$ 77 $\pm$ 71\\
			$q5$ & 497 $\pm$ 66 $\pm$ 60 & 775 $\pm$ 84 $\pm$ 75\\
			$q6$ & 613 $\pm$ 73 $\pm$ 227 & 809 $\pm$ 86 $\pm$ 164\\
			\hline
			\hline
		\end{tabular}
		\label{tab:sig_yields}
	\end{center} 
\end{table}

\section{Systematic uncertainties}
\label{sec:sys}
The fractional uncertainties on the corrected signal yields in each $q^{2}$ bin from various sources of systematic uncertainty are shown in Table~\ref{tab:sig_systematics}.
All systematic uncertainties are evaluated using the same approach.
For each source of uncertainty, we vary the templates 1000 times by sampling from Gaussian distributions of the central values. 
For example, to evaluate the uncertainties due to the \Btorholnu{} form factors, we sample 1000 alternative \BtoXulnu{} distributions by assuming the form-factor parameter uncertainties follow Gaussian distributions. 
We create 1000 simplified simulated data (toy) distributions by adding the resulting variations to any unaffected templates.
We then fit the nominal templates to the toy distributions and obtain a covariance matrix for each source of uncertainty using Pearson correlation~\cite{benesty2009pearson}.

The largest contribution to the uncertainty in the \BtoXulnu{} background template comes from the uncertainty in the \Btorholnu{} form-factor parameters.
The effect of the \Btorholnu{} form-factor and branching fraction uncertainties is included in the \Btorholnu{} category in Table~\ref{tab:sig_systematics}.
We also evaluate the uncertainties due to the \Btopiellnu{}, $B\rightarrow \omega \ell\nu$, $B\rightarrow \eta \ell\nu$, and $B\rightarrow \eta' \ell\nu$ form factors, and obtain uncertainties shown under the \BtoXulnu{} category.
This category also includes the effects of uncertainties of the exclusive and inclusive \BtoXulnu{} branching fractions, except for the $B\rightarrow \rho \ell\nu$ branching fraction.
The \BtoXclnu{} category in Table~\ref{tab:sig_systematics} includes the effects of the uncertainties of the \BtoDlnu{} and \BtoDstlnu{} form-factor parameters, and the exclusive and inclusive \BtoXclnu{} branching fractions.

The detector uncertainties include uncertainties arising from the tracking efficiency and the corrections to the lepton- and pion-identification efficiencies.
The effect of having a small sample of simulated data is also considered.

Two sources of uncertainty are included in the continuum category in Table~\ref{tab:sig_systematics}. 
One is the uncertainty on the signal yields due to the limited off-resonance sample size.
The resulting uncertainties range from 3\% to 11\% in the intermediate $q^{2}$ bins and dominate the uncertainties in the highest and lowest $q^{2}$ bins, in which the continuum background is largest.
The other source of uncertainty originates from the assumption that the continuum reweighting in $q^{2}$ bins is independent of \deltae{} and \mbc{}.
To estimate this, we parameterize the shape difference between simulated continuum and off-resonance data as a linear function in \deltae{}.
We fit this parameterization to the observed difference and obtain central values and a covariance for the two parameters.
We then follow the same procedure used in the evaluation of the other systematic uncertainties.
We sample 1000 varied continuum templates, generate the corresponding toy distributions, and by fitting the nominal templates to the toy distributions obtain refitted yields.
If the difference between the nominal fit central value and the mean value of the refitted yields is significant, we take this difference as an additional uncertainty on the yield.
This introduces large uncertainties to the highest and lowest $q^{2}$ bins in the electron and muon modes, respectively.

We consider additional systematic uncertainties from the number of $B\overline{B}$ pairs $N_{B\overline{B}}$, and the branching fraction of $\Upsilon(4S)\rightarrow B^{0}\overline{B^{0}}$, $f = 0.486 \pm 0.006$~\cite{Zyla:2020zbs}.
These uncertainties do not affect the yields, but they contribute a 2\% systematic uncertainty to the branching fraction results.

\begin{table*}[ht!]
	\begin{center}
	\caption{Summary of fractional uncertainties on the yields.}
    \begin{tabular}{ccccccccccccc}
        \hline
        \hline
    	Source &\multicolumn{6}{c}{\Btopienu} &\multicolumn{6}{c}{\Btopimunu}\\
    	&$q1$ & $q2$ & $q3$ & $q4$& $q5$ & $q6$ &$q1$ & $q2$ & $q3$ & $q4$& $q5$ & $q6$  \\
    	\hline
    	Detector &1.2&1.0&1.1&1.4&2.3&2.4&  2.3&3.2&3.3&1.2&1.9&3.8\\
    	MC sample size & 4.0& 2.0& 2.4 & 2.8 & 3.9&5.6& 3.9 & 2.0& 2.3 & 2.7 & 3.4&4.8 \\
    	Continuum & 13.1& 5.5& 4.4 & 7.8 & 10.5 & 33.9  & 53.3& 8.8& 3.2 & 4.5& 8.0&11.4 \\
    	\hline
    	$B\rightarrow \rho\ell\nu$ &9.5&12.5&9.7&6.9&3.4&12.9&8.7&11.6&8.6&6.3&3.3&14.3 \\
    	$B\rightarrow X_{u}\ell\nu$ &3.3&1.9&2.1&2.1&1.8&3.7& 3.4&2.3&2.0&2.3&2.1&6.0\\
        $B\rightarrow X_{c}\ell\nu$  &2.3&3.0&1.1&0.8&0.5&2.4&2.4&1.5&1.5&0.8&0.5&2.2\\
    	\hline
    	Total syst. &17.2&14.3& 11.2& 11.1&12.0 & 37.0&53.4&15.2& 10.3& 8.7&9.7&20.3 \\
    	Stat.& 10.2 & 6.01 & 6.86 & 8.08& 10.3& 13.2& 10.4& 6.0& 6.4 & 7.8 & 9.7 & 13.4 \\
    	\hline	
    	Total &20.2&15.5& 13.2& 13.7&15.9&39.2& 54.5& 16.4& 12.2& 11.6&13.7&24.3\\
    	\hline
    	\hline
    \end{tabular}
	\label{tab:sig_systematics}
	\end{center} 
\end{table*}

\section{Results}
\subsection{Branching fractions}
The partial branching fraction in $q^{2}$ bin $i$ is calculated using the corrected yield, $N_{i}$, and the corresponding signal efficiency, $\epsilon_{i}$, from
\begin{equation}
	\Delta \mathcal{B}_{i} = \frac{N_{i}} {2 f \epsilon_{i}\, N_{B\overline{B}}}.
	\label{eq:deltaBF}
\end{equation} 
The results for \Btopienu{} and \Btopimunu{} decays, $\Delta \mathcal{B}_{i,\mathrm{e}}$ and $\Delta \mathcal{B}_{i,\mathrm{\mu}}$, respectively, are given in Table~\ref{tab:Full_BF}. 
We also provide an average over both decay channels, fully correlating common systematic uncertainties in Table~\ref{tab:Full_BF}\footnotemark[1].
The total covariance matrix of the averaged partial branching fractions is given in Table~\ref{tab:BF_cov_total}.
The total branching fraction determined from the sum of the averaged partial branching fractions is
\begin{center}
	$\mathcal{B}(\Bztopiellnu) = (1.426 \pm 0.056 (\mathrm{stat}) \pm 0.125 (\mathrm{syst})) \times 10^{-4}$,\footnotemark[1]
\end{center}
where the first uncertainty is statistical and the second is systematic.

\begin{table}[ht!]
	\begin{center}
	\caption{$\Delta \mathcal{B}$ ($\times 10^{4}$) in each $q^{2}$ bin calculated from the corrected signal yields for the \Btopienu{} and \Btopimunu{} modes. The averaged partial branching fractions for \Bztopiellnu{} are also shown. The first uncertainty is statistical and the second is systematic.}
		\begin{tabular}{cccc}
		    \hline
		    \hline
			$q^{2}$ bin & \multicolumn{1}{c}{\Btopienu} & \multicolumn{1}{c}{\Btopimunu}& \multicolumn{1}{c}{\Bztopiellnu}\\
			\hline
			$q1$ &0.261 $\pm$ 0.030 $\pm$ 0.045 &0.281 $\pm$ 0.028 $\pm$ 0.150 &0.272 $\pm$ 0.031 $\pm$ 0.044\\
			$q2$ &0.298 $\pm$ 0.018 $\pm$ 0.043 &0.285 $\pm$ 0.016 $\pm$ 0.044 &0.290 $\pm$ 0.013 $\pm$ 0.040\\
			$q3$ &0.268 $\pm$ 0.020 $\pm$ 0.030 &0.290 $\pm$ 0.018 $\pm$ 0.030 &0.279 $\pm$ 0.014 $\pm$ 0.028 \\
			$q4$ &0.196 $\pm$ 0.022 $\pm$ 0.022 &0.204 $\pm$ 0.019 $\pm$ 0.018 &0.199 $\pm$ 0.015 $\pm$ 0.017\\
			$q5$ &0.180 $\pm$ 0.024 $\pm$ 0.022 &0.194 $\pm$ 0.021 $\pm$ 0.019 &0.188 $\pm$ 0.016 $\pm$ 0.015 \\
			$q6$ &0.200 $\pm$ 0.024 $\pm$ 0.074 &0.220 $\pm$ 0.023 $\pm$ 0.045 &0.198 $\pm$ 0.031 $\pm$ 0.042 \\
			\hline
			\hline
		\end{tabular}
		\label{tab:Full_BF}
	\end{center} 
\end{table}

\footnotetext{Note: these results differ slightly from those presented at the 2022 conferences.}

\begin{table}[ht!]
	\begin{center}
	\caption{Total covariance matrix of the averaged partial branching fractions $\Delta \mathcal{B}$ for \Bztopiellnu{} in units of $10^{-13}$ .}
		\begin{tabular}{ccccccc}
		    \hline
		    \hline
			$q^{2}$ bin&$q1$ & $q2$ & $q3$ & $q4$ & $q5$& $q6$\\
			\hline
			$q1$&  \, 2.913 & 0.952 & 0.663 & 0.290 & 0.039 & -0.166 \\
			$q2$ & \, & 1.753 & 0.985 & 0.483 & 0.055 & 0.216 \\
			$q3$ & \, && 0.983 & 0.403 & 0.081 & 0.077 \\
			$q4$ & \, &&& 0.503 & 0.122& 0.120 \\
			$q5$ & \, &&&& 0.488 & 0.364 \\
			$q6$ & \, &&&&& 2.697 \\
			\hline
			\hline
		\end{tabular}
		\label{tab:BF_cov_total}
	\end{center} 
\end{table}

\subsection{$|V_{ub}|$ determination}
We extract $|V_{ub}|$ using $\chi^{2}$ fits to the measured $q^{2}$ spectra.
We include lattice QCD constraints on the eight BCL parameters from Ref.~\cite{FermilabMILC} as nuisance parameters.
These constrain the shape and normalization of the relevant form factors entering the differential decay rate and allow for a determination of $|V_{ub}|$. 

The $\chi^{2}$ is defined as
\begin{equation}
	\chi^{2} = \sum_{i,j = 1}^{6}(\Delta \mathcal{B}_{i}- \Delta \Gamma_{i}\tau) C^{-1}_{ij} (\Delta \mathcal{B}_{j} - \Delta \Gamma_{j}\tau) + \chi^{2}_{\mathrm{LQCD}},
	\label{eq:minim}
\end{equation} 
where $C^{-1}_{ij}$ is the inverse total covariance matrix of the measured partial branching fractions in bin $i$, $\mathcal{B}_{i}$.
The quantities $\Delta\Gamma_{i}$ contain the predictions for the partial decay rates in bin $i$, $\tau$ is the $B^{0}$ lifetime, and $\chi^{2}_{\mathrm{LQCD}}$ incorporates the constraints from the lattice calculation.

The obtained results are
\begin{center}
	$|V_{ub}|_{\Btopienu} = (3.60 \pm 0.18 (\mathrm{stat}) \pm 0.14 (\mathrm{syst}) \pm 0.18 (\mathrm{theo})) \times 10^{-3}$\\
	$|V_{ub}|_{\Btopimunu} = (3.71 \pm 0.16 (\mathrm{stat}) \pm 0.15 (\mathrm{syst}) \pm 0.17 (\mathrm{theo})) \times 10^{-3}$,
	$|V_{ub}|_{\Bztopiellnu} = (3.55 \pm 0.12 (\mathrm{stat}) \pm 0.13 (\mathrm{syst}) \pm 0.17 (\mathrm{theo})) \times 10^{-3}$.\footnotemark[1]
\end{center} 

\begin{figure*}[ht!]
	\centering
		\includegraphics[width=0.85\linewidth]{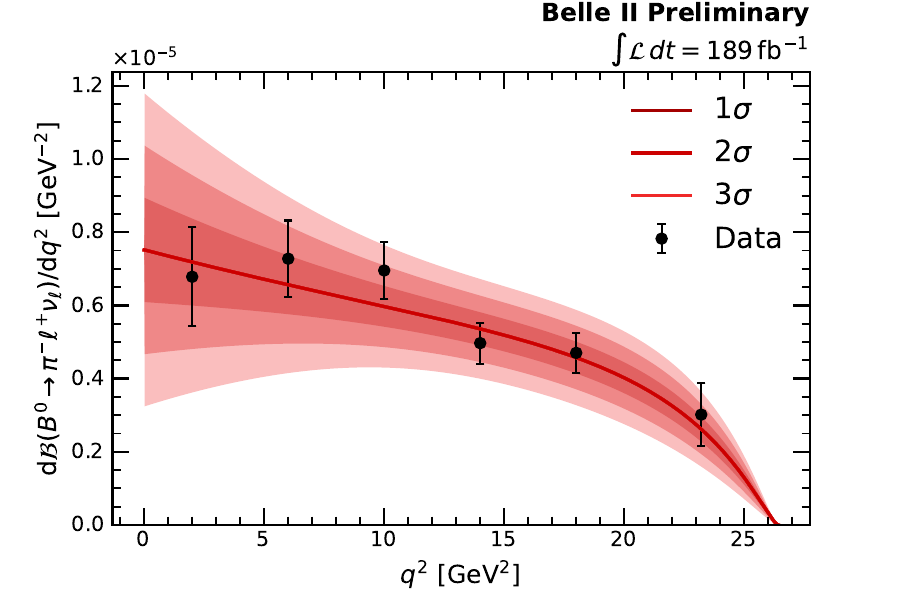}
	\caption{Measured averaged partial branching fractions as a function of $q^{2}$. The fitted differential rate is shown together with the one, two, and three standard-deviation uncertainty bands.}
	\label{fig:Vub}
\end{figure*}

The value of $|V_{ub}|$ determined by fitting the averaged \Bztopiellnu{} partial branching fractions is lower than the $|V_{ub}|$ results from the \Btopienu{} and \Btopimunu{} samples.
This feature of the fit arises because the extraction of $|V_{ub}|$ is most sensitive to the high $q^{2}$ region, where the average partial branching fraction listed in Table~\ref{tab:Full_BF} is also lower than the \Btopienu{} and \Btopimunu{} partial branching fractions due to correlations between $q^2$ bins.
The partial branching fractions of \Bztopiellnu{} measured as a function of $q^{2}$ are shown in Figure~\ref{fig:Vub}. 
The fitted differential rate is also shown, and the one, two, and three standard-deviation uncertainty bands are given.

\section{Summary}
We extract partial branching fractions for \Btopienu{} and \Btopimunu{} decays reconstructed in an electron-positron collision data sample corresponding to 189~fb$^{-1}$ collected by the Belle~II experiment in 2019-2021.
By averaging the results we obtain partial branching fractions for \Bztopiellnu{}.
The total branching fraction result of \Bztopiellnu{} is found to be $(1.426 \pm 0.056(\mathrm{stat}) \pm 0.125 (\mathrm{syst})) \times 10^{-4}$.
This is consistent with the current world average of $(1.50 \pm 0.06) \times 10^{-4}$~\cite{Zyla:2020zbs}.
Currently our results are limited by the size of the off-resonance data set.
This uncertainty could be reduced by improvements in the simulation of continuum backgrounds.
We also extract values of the CKM matrix-element magnitude $|V_{ub}|$ and from \Bztopiellnu{} decays obtain $(3.55 \pm 0.12 (\mathrm{stat})\pm 0.13 (\mathrm{syst}) \pm 0.17 (\mathrm{theo})) \times 10^{-3}$.
This agrees with the current value obtained by HFLAV~\cite{Amhis:2019ckw} of $(3.70 \pm 0.10 (\mathrm{exp}) \pm 0.12 (\mathrm{theo})) \times 10^{-3}$.

%% file: acknowledgements.tex
We thank the SuperKEKB group for the excellent operation of the
accelerator; the KEK cryogenics group for the efficient
operation of the solenoid; the KEK computer group for
on-site computing support; and the raw-data centers at
BNL, DESY, GridKa, IN2P3, and INFN for off-site computing support.